\UseRawInputEncoding

\documentclass[10pt, prb, aps, twocolumn, showpacs, citeautoscript, floatfix, reprint, amsmath, amssymb, notitlepage, superscriptaddress]{revtex4-1}

\usepackage{graphicx}
\usepackage{stmaryrd}
\usepackage{rotating}
\usepackage{amsmath}
\usepackage{amsfonts}
\usepackage{amssymb}
\usepackage{wasysym}
\usepackage[countmax]{subfloat}
\usepackage{dcolumn} 
\usepackage{bm} 
\usepackage{color}

\usepackage{float}
\usepackage{latexsym}

\begin{document}

\title{Opacity of graphene independent of light frequency and polarization due to the topological charge of the Dirac points}

\author{Matheus S. M. de Sousa}

\affiliation{Department of Physics, PUC-Rio, 22451-900 Rio de Janeiro, Brazil}

\author{Wei Chen}

\affiliation{Department of Physics, PUC-Rio, 22451-900 Rio de Janeiro, Brazil}

\date{\rm\today}

\begin{abstract}

The opacity of graphene is known to be approximately given by the fine-structure constant $\alpha$ times $\pi$. We point out the fact that the opacity is roughly independent of the frequency and polarization of the light can be attributed to the topological charge of the Dirac points. As a result, one can literally see the topological charge by naked eyes from the opacity of graphene, and moreover it implies that the fine-structure constant is topologically protected. A similar analysis suggests that 3D topological insulator thin films of any thickness also have opacity $\pi\alpha$ in the infrared region owing to the topological surface states, indicating that one can see the surface states by naked eyes through an infrared lens. For 3D Dirac or Weyl semimetals, the optical absorption power is linear to the frequency in the infrared region, with a linearity given by the fine-structure constant and the topological charge of Weyl points.

\end{abstract}

\maketitle

\section{Introduction}

A fundamentally important issue in the research of topological materials is how the topological order can be unambiguously detected in experiments. Conventionally, the most straightforward way is to detect the effects related to the metallic edge or surface states, such as the quantized Hall conductance\cite{vonKlitzing80,Konig07} or zero bias conductance\cite{Mourik12}, which only occur in topologically nontrivial phase and is directly determined by the bulk topological invariant. On the other hand, to our knowledge, there has yet been a bulk material property in the macroscopic scale that can be felt by human perception and be attributed to the topological order. This is in sharp contrast to the materials that possess Landau order parameters. For instance, magnetic order can be simply perceived from the force that repels or attracts two bar magnets, and superconductivity can be understood as the mechanism behind the magnetic levitation.

In this paper, we point out that the topological charge of various materials containing gapless Dirac cones can actually be seen by naked eyes, either directly or through an infrared lens. This statement is made based on our discovery that for gapless Dirac materials, the optical absorption power is always proportional to the fine-structure constant $\alpha=e^{2}/4\pi\hbar c\varepsilon_{0}\approx 1/137$ times a factor determined by the topological charge ${\cal C}$. This feature originates from a metric-curvature correspondence between topological order and quantum metric of the valence bands\cite{vonGersdorff21_metric_curvature}, and moreover the later is directly measurable by optical absorption power. A particularly important application of our theory is graphene, since single layer graphene of size up to cm$^{2}$ deposit on transparent substrates, such as polyethylene terephthalate (PET) or quartz, is already commercially available. In this case, the well-known $\pi\alpha\approx 2.3\%$ opacity of graphene can be easily perceived by naked eyes\cite{Nair08,Stauber08,Bruna09,Weber10}. Remarkably, despite this $\pi\alpha$ opacity has been well-known for more than a decade, it has never been pointed out until the present work that the approximate frequency and polarization independence of the opacity $\pi\alpha\times 4{\cal C}^{2}$ is due to the protection by the topological charge ${\cal C}=1/2$. Similar to the seminal Thouless-Kohmoto-Nightingale-den Nijs (TKNN) theory that links the DC Hall conductivity to the Chern number\cite{Thouless82}, we prove this feature by utilizing a linear response theory to link the longitudinal optical conductivity to the topological charge. On the other hand, we also investigate how several realistic factors in graphene, such as the hexagonal warping and van Hove singularity, render the opacity not exactly constant of frequency and polarization, and discuss the possibility of extracting the fine-structure constant $\alpha$ precisely from the opacity. Furthermore, through the inclusion of impurities, our theory well explains the reduction of opacity in fluorinated graphene at low frequency\cite{Nair10}.

Another remarkable prediction of our theory is that one can literally see the topological surface states of 3D topological insulator (TI) thin films by naked eyes, since they are also described by gapless 2D Dirac cones. Our prediction is that all single crystal thin films of 3D TIs, such as\cite{Zhang09,Liu10} Bi$_{2}$Se$_{3}$ and Sb$_{2}$Te$_{3}$, have the same $\pi\alpha\times 4{\cal C}^{2}\approx 2.3\%$ opacity as graphene in the infrared region regardless the thickness of the film. This feature can be verified by simply looking at 3D TI thin films of different thickness through an infrared lens, wich should all show the same opacity, offering a very pedagogical way to perceive the surface states. Finally, we turn to 3D Weyl and Dirac semimetals to elaborate that the well-known optical absorption power that is linear in frequency in the infrared region is also topologically protected\cite{Hosur12,Bacsi13,Timusk13,Ashby14,Xu16}. Thus the fact that, through an infrared lens, these semimetals look darker under higher frequency light is also a topological phenomenon.



\section{Optical absorption of gapless Dirac models as a topological charge}

\subsection{Relating optical absorption power to quantum geometry \label{sec:relating_absorption_to_quantum_geometry}} 

Our survey starts by considering the quantum geometry of valance band states\cite{Matsuura10,vonGersdorff21_metric_curvature,Chen22_dressed_Berry_metric}. We will reserve the index $n$ for valence bands, $m$ for conduction bands, $\ell$ for all the bands, and their energies at momentum ${\bf k}$ are denoted by $\varepsilon_{\ell}^{\bf k}$. For a gaped topological material with $N_{-}$ valence bands, the fully antisymmetric valence band Bloch state at momentum ${\bf k}$ is $|u^{\rm val}({\bf k})\rangle=\epsilon^{n_{1}n_{2}...n_{N-}}|n_{1}\rangle|n_{2}\rangle...|n_{N_{-}}\rangle/\sqrt{N_{-}!}$. 
The quantum metric of this state is defined from\cite{Provost80} 
\begin{eqnarray}
|\langle u^{\rm val}({\bf k})|u^{\rm val}({\bf k+\delta k})\rangle|=1-\frac{1}{2}g_{\mu\nu}({\bf k})\delta k^{\mu}\delta k^{\nu},
\label{uval_gmunu}
\end{eqnarray}
whose diagonal component $g_{\mu\mu}$ can be calculated from each band by\cite{vonGersdorff21_metric_curvature} 
\begin{eqnarray}
&&g_{\mu\mu}({\bf k})=\langle \partial_{\mu}u^{\rm val}|\partial_{\mu}u^{\rm val}\rangle
-\langle \partial_{\mu}u^{\rm val}|u^{\rm val}\rangle \langle u^{\rm val}|\partial_{\mu}u^{\rm val}\rangle
\nonumber \\
&&=\sum_{nm}\langle \partial_{\mu}n|m\rangle\langle m|\partial_{\mu}n\rangle.
\label{gmunu_T0}
\end{eqnarray}
To relate the quantum metric to optical responses, we introduce a quantum metric spectral function\cite{Chen22_dressed_Berry_metric} 
\begin{eqnarray}
&&g_{\mu\mu}({\bf k},\omega)=\sum_{\ell<\ell '}\langle\partial_{\mu}\ell|\ell '\rangle\langle\ell '|\partial_{\mu}\ell\rangle
\nonumber \\
&&\times\left[f(\varepsilon_{\ell}^{\bf k})-f(\varepsilon_{\ell '}^{\bf k})\right]\delta(\omega+\frac{\varepsilon_{\ell}^{\bf k}}{\hbar}-\frac{\varepsilon_{\ell '}^{\bf k}}{\hbar})
\nonumber \\
&&=\frac{V_{D}}{\pi e^{2}\hbar\omega}\sigma_{\mu\mu}({\bf k},\omega),
\label{gmunuw_finiteT_general}
\end{eqnarray}
which at zero temperature frequency-integrates to the quantum metric $\lim_{T\rightarrow 0}\int_{0}^{\infty}d\omega\,g_{\mu\mu}({\bf k},\omega)=g_{\mu\mu}({\bf k})$, where $V_{D}$ is the volume of the $D$-dimensional unit cell, and $\sigma_{\mu\mu}({\bf k},\omega)$ is the finite temperature longitudinal optical conductivity at momentum ${\bf k}$ obtained from linear response theory\cite{deSousa23_fidelity_marker}.

The optical conductivity measured in real space is given by the momentum integration
\begin{eqnarray}
&&\sigma_{\mu\mu}(\omega)=V_{D}\int\frac{d^{D}{\bf k}}{(2\pi\hbar)^{D}}\,\sigma_{\mu\mu}({\bf k},\omega)
\nonumber \\
&&=\frac{\pi e^{2}}{\hbar^{D-1}}\,\omega\int\frac{d^{D}{\bf k}}{(2\pi)^{D}}\,g_{\mu\mu}({\bf k},\omega).
\label{fidelity_number_spec_fn}
\end{eqnarray}
Furthermore, applying an oscillating electric field polarized in $\mu$ direction $E_{\mu}(\omega,t)=E_{0}\cos\omega t$ to the system induces a current that oscillates accordingly $j_{\mu}(\omega,t)=\sigma_{\mu\mu}(\omega)E_{0}\cos\omega t$, where $E_{0}$ is the strength of the field. Thus the optical absorption power per unit cell at frequency $\omega$ is\cite{deSousa23_fidelity_marker} 
\begin{eqnarray}
&&W_{a}^{\mu}(\omega)=\langle j_{\mu}(\omega,t)E_{\mu}(\omega,t)\rangle_{t}=\frac{1}{2}\sigma_{\mu\mu}(\omega)E_{0}^{2},
\label{absorption_power_global}
\end{eqnarray}
where the time average gives $\langle\cos^{2}\omega t\rangle_{t}=1/2$. The main point of the present work is how $W_{a}^{\mu}(\omega)$ is related to the topological charge ${\cal C}$ and fine-structure constant $\alpha$ in a topologically protected manner, as we elaborate below for several different topological materials.


\subsection{Opacity of pristine graphene \label{sec:opacity_pristine_graphene}} 

The low energy band structure of graphene can be described by the tight-binding model on a honeycomb lattice with nearest-neighbor hopping $H=\sum_{\langle ij\rangle\sigma}t\,c_{i\sigma}^{\dag}c_{j\sigma}$ with $t\approx 2.8$eV, and we denote the distance between neighboring carbon atoms by $a=0.142$nm\cite{CastroNeto09}. For each spin species, the low energy Homiltonian in the momentum space may be obtained by an expansion around the two Dirac points ${\bf K}$ and ${\bf K}'$, yielding the linear Dirac Hamiltonian \cite{Bernevig13}
\begin{eqnarray}
&&H_{0}^{{\bf K},{\bf K'}}({\bf k})=v_{F}\left(\pm k_{y}\sigma_{x}-k_{x}\sigma_{y}\right),
\label{graphene_linear_Dirac_Hamiltonian}
\end{eqnarray}
where $v_{F}=3ta/2\hbar$ is the Fermi velocity. This linearized model well describes the linear band structure up to energy $\sim 1$eV, and hence is a suitable model for the optical absorption in the visible light range. To proceed, we introduce the spin-valley index $\gamma=\left\{{\bf K}\uparrow,{\bf K}\downarrow,{\bf K}'\uparrow,{\bf K}'\downarrow\right\}$ and denote the valence and conduction band states by $|n^{\gamma}\rangle$ and $|m^{\gamma}\rangle$, whose eigenenergies $\varepsilon_{n}^{\bf k}=-v_{F}k$ and $\varepsilon_{m}^{\bf k}=v_{F}k$ do not depend on $\gamma$. The topological charge ${\cal C}$ per spin at each of the two Dirac points ${\bf K}$ and ${\bf K'}$ is given by integrating the valence band Berry connection along a closed loop of radius $k$ circulating the Dirac points
\begin{eqnarray}
&&\oint \frac{d\phi}{2\pi}\langle n^{\bf K\uparrow}|i\partial_{\phi}|n^{\bf K\uparrow}\rangle=-\oint \frac{d\phi}{2\pi}\langle n^{\bf K'\uparrow}|i\partial_{\phi}|n^{\bf K'\uparrow}\rangle
\nonumber \\
&&=-1/2\equiv-{\cal C},
\label{eq:grapheneC}
\end{eqnarray}
which has opposite signs at the two Dirac points. 


We now elaborate the relation between this topological charge and the quantum metric. The quantum metric for a valence band state of spin-valley flavor $\gamma$ is defined from the overlap $|\langle n^{\gamma}({\bf k})|n^{\gamma}({\bf k+\delta k})\rangle|=1-g_{\mu\nu}^{\gamma}\delta k^{\mu}\delta k^{\nu}/2$. In particular, because the valence band state $|n^{\gamma}\rangle$ does not depend on the module of the momentum $k$, the only nonzero component of the quantum metric in the polar coordinates $\left\{\mu,\nu\right\}=\left\{k,\phi\right\}$ is the azimuthal component, which turns out to be equal to the square of the topological charge
\begin{eqnarray}
&&g_{\phi\phi}^{\gamma}=|\langle m^{\gamma}|i\partial_{\phi}|n^{\gamma}\rangle|^{2}=|\langle n^{\gamma}|i\partial_{\phi}|n^{\gamma}\rangle|^{2}={\cal C}^{2}=\frac{1}{4}.\;\;\;
\label{gphiphi}
\end{eqnarray}
This relation, which has been called the metric-curvature correspondence\cite{vonGersdorff21_metric_curvature}, is the key to identify the opacity of graphene with the topological charge, as we shall see below. 

The quantum metric in the Cartesian coordinates $\left\{\mu,\nu\right\}=\left\{x,y\right\}$ can be obtained from Eq.~(\ref{gphiphi}) by utilizing $\partial_{x}=\cos\phi\,\partial_{k}-(\sin\phi/k)\partial_{\phi}$ and $\partial_{y}=\sin\phi\,\partial_{k}+(\cos\phi/k)\partial_{\phi}$.
Moreover, because in the angular integration $\int_{0}^{2\pi}d\phi$ in the calculation of optical absorption below, the choice of $\phi=0$ is arbitrary, so for any polarization $\mu=\left\{x,y\right\}$ we can simply set 
\begin{eqnarray}
g_{\mu\mu}^{\gamma}=\frac{\sin^{2}\phi}{k^{2}}g_{\phi\phi}^{\gamma}=\frac{\sin^{2}\phi}{k^{2}}\,{\cal C}^{2}.
\label{gmumu_graphene}
\end{eqnarray} 
It follows that the quantum metric spectral function in Eq.~(\ref{gmunuw_finiteT_general}) for the flavor $\gamma$ is
\begin{eqnarray}
&&g_{\mu\mu}^{\gamma}({\bf k},\omega)=g_{\mu\mu}^{\gamma}
\left[f(\varepsilon_{n}^{\bf k})-f(\varepsilon_{m}^{\bf k})\right]\delta(\omega+\frac{\varepsilon_{n}^{\bf k}}{\hbar}-\frac{\varepsilon_{m}^{\bf k}}{\hbar}).\;\;\;
\label{gmunuw_finiteT}
\end{eqnarray}
In addition, from Eq.~(\ref{gmunuw_finiteT_general}), the optical conductivity contributed by the flavor $\gamma$ is $\sigma_{\mu\mu}^{\gamma}({\bf k},\omega)=\pi e^{2}\hbar\omega\,g_{\mu\mu}^{\gamma}({\bf k},\omega)/A_{cell}$,
where $A_{cell}=3\sqrt{3}a^{2}/2$ is the unit cell area of the hexagonal lattice. Putting the spectral function in Eq.~(\ref{gmunuw_finiteT}) into Eq.~(\ref{fidelity_number_spec_fn}), summing over $\gamma$, and using the area of the BZ $A_{BZ}=8\pi^{2}/3\sqrt{3}a^{2}$, the conductivity measured in real space is 
\begin{eqnarray}
\sigma_{\mu\mu}(\omega)=\frac{e^{2}}{\hbar}{\cal C}^{2}\left[f\left(-\frac{\hbar\omega}{2}\right)-f\left(\frac{\hbar\omega}{2}\right)\right].
\end{eqnarray}
indicating that the optical conductivity\cite{Falkovsky07,Falkovsky08} is given by the conductance quantum times the topological charge.


For an incident light with electric field $E_{0}$, the incident power per unit cell of area $A_{cell}$ is $W_{i}=c\varepsilon_{0}E_{0}^{2}/2$. Using Eq.~(\ref{absorption_power_global}), the opacity at polarization $\mu$ and frequency $\omega$ is 
\begin{eqnarray}
&&{\cal O}(\omega)=\frac{W_{a}^{\mu}(\omega)}{W_{i}}
\nonumber \\
&&=\pi\alpha\times 4{\cal C}^{2}\left[f\left(-\frac{\hbar\omega}{2}\right)-f\left(\frac{\hbar\omega}{2}\right)\right],
\label{Ow_graphene_finiteT}
\end{eqnarray}
which does not depend on the polarization $\mu$. In the zero temperature limit, one obtains 
\begin{eqnarray}
\lim_{T\rightarrow 0}{\cal O}(\omega)=\pi\alpha\times 4{\cal C}^{2}=\pi\alpha\approx 2.3\%.
\end{eqnarray}
This is the seminal result of the $2.3\%$ opacity (or transmittance) of graphene\cite{Nair08,Stauber08,Bruna09,Weber10}, and our analysis brings in several new aspects to this result: (1) The independence of frequency $\omega$ is due to the topological charge ${\cal C}=1/2$, or equivalently the azimuthal quantum metric $g_{\phi\phi}^{\gamma}$ according to Eq.~(\ref{gphiphi}), that is independent of the circle of radius $k=\hbar\omega/2v_{F}$ at which the conduction band electrons are excited by the light. In addition, $g_{\phi\phi}^{\gamma}$ is also independent of the angle $\phi$, hence the opacity does not depend on the polarization $\mu$ either. Since the visible light produced by common light sources usually contains a wide range of polarization and frequency, the reason that graphene always shows $2.3\%$ opacity is due to the topological protection of the Dirac cone, and it implies that one can literally see the topological charge ${\cal C}$ by naked eyes through the opacity. (2) One may use the plateau of frequency or polarization dependence of opacity to extract $\alpha$. This procedure echoes the extraction of von Klitzing constant $h/e^{2}$ from the Hall plateaux as a function of magnetic field in the quantum Hall effect (QHE)\cite{vonKlitzing80}, which is also topologically protected. Our work that recognizes the opacity of graphene as a topological charge through a linear response theory of longitudinal optical conductivity is thus conceptually analogous to the seminal TKNN theory, which recognizes the quantized Hall conductance as a topological invariant through a linear response theory of the DC Hall conductance\cite{Thouless82}. This topological protection implies that the measured $\alpha$ should be independent of many details of the system, i.e., ideally any 2D materials that have a Dirac code should have the same $\pi\alpha$ opacity, which may describe a variety of 2D materials such as graphynes\cite{Malko12,Malko12_2}, B$_{2}$S\cite{Zhao18_B2S}, silicene\cite{Ezawa13,Houssa15,Sadeddine17}, germanene\cite{Acun15}, etc\cite{Wehling14,Balendhran15,Wang15_Dirac_cone_materials_review}.






\subsection{Influence of realistic factors on the opacity of graphene \label{sec:realistic_factors_opacity}}

\begin{figure}[ht]
\centering
\includegraphics[width=0.99\linewidth]{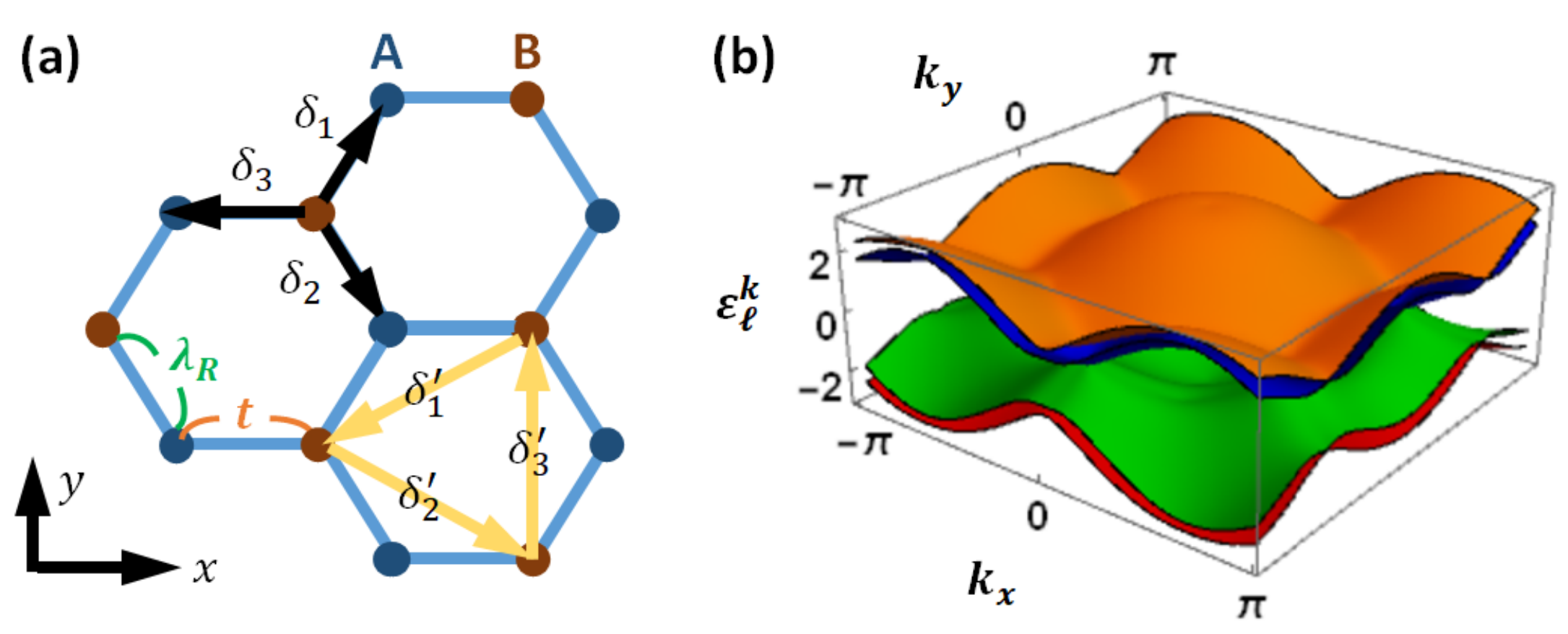}
\caption{(a) The definition of coordinates and various vectors on the honeycomb lattice. (b) The band structure simulated by $t=1$, $t'=0.036$, and $\lambda_{R}=0.2$. }
\label{fig:graphene_RSOC_band}
\end{figure}

Despite the appealing connection between the opacity and topological charge, the above simple results suffer a great challenge from many realistic factors in 2D materials containing Dirac cones, as we formulate below. The first is the realistic band structure beyond the simple description of Eq.~(\ref{graphene_linear_Dirac_Hamiltonian}), which contains complications such as hexagonal warping, next-nearest-neighbor hopping, and Rashba spin-orbit coupling (RSOC). To investigate their effects, we incorporate them into the tight-binding model defined on a honeycomb lattice 
\begin{eqnarray}
H&=&-t\sum_{\langle ij\rangle,\sigma}c_{i\sigma}^{\dag}c_{j\sigma}+t'\sum_{\langle\langle ij\rangle\rangle,\sigma}c_{i\sigma}^{\dag}c_{j\sigma}
\nonumber \\
&+&i\lambda_{R}\sum_{\langle ij\rangle,\alpha,\beta}c_{i\alpha}^{\dag}\left({\boldsymbol \sigma}_{\alpha\beta}\times{\bf d}_{ij}\right)^{z}c_{j\beta}.
\label{Hamiltonian_graphene_Rashba_mag}
\end{eqnarray}
Here $ c^{\dag}_{i \sigma} (c_{i \sigma}) $ creates (annihilates) an electron of spin $ \sigma $ on the lattice site $ i$, and $ \langle i j \rangle $ and $ \langle\langle i j \rangle\rangle $ indicate nearest-neighbor and next-nearest-neighbor lattice sites with the corresponding hopping $ - t $ and $t$. The $\lambda_{R}$ is the RSOC coupling constant, ${\boldsymbol \sigma}=(\sigma^{x},\sigma^{y},\sigma^{z})$ are the spin Pauli matrices, ${\bf d}_{ij}$ is the vector connecting the site $i$ to $j$. Defining the basis $\psi=\left(A\uparrow,B\uparrow,A\downarrow,B\downarrow\right)$ and the Fourier transformation $c_{Ii\sigma}=\sum_{\bf k}e^{i{\bf k\cdot r}_{i}}c_{I{\bf k}\sigma}$, where $I=\left\{A,B\right\}$ denotes the two sublattices and ${\bf r}_{i}$ the unit cell position, the Hamiltonian $H=\sum_{\bf k IJ\alpha\beta}c_{I{\bf k}\alpha}^{\dag}H_{I\alpha J\beta}({\bf k})c_{J{\bf k}\beta}$ is described by the matrix
\begin{eqnarray}
&&H_{I\alpha J\beta}({\bf k})=
\left(
\begin{array}{cccc}
t'Z' & tZ^{\ast} & 0 & \lambda_{R}Y^{\ast} \\
tZ & t'Z' & \lambda_{R}X^{\ast} & 0 \\
0 & \lambda_{R}X & t'Z' & tZ^{\ast} \\
\lambda_{R}Y & 0 & tZ & t'Z'
\end{array}
\right),\;\;\;\;
\nonumber \\
&&Z\equiv e_{1}+e_{2}+e_{3},
\nonumber \\
&&Z'\equiv e'_{1}+e'_{2}+e'_{3},
\nonumber \\
&&X\equiv\frac{-1-i\sqrt{3}}{2}e_{1}^{\ast}+\frac{-1+i\sqrt{3}}{2}e_{2}^{\ast}+e_{3}^{\ast},
\nonumber \\
&&Y\equiv\frac{1+i\sqrt{3}}{2}e_{1}+\frac{1-i\sqrt{3}}{2}e_{2}-e_{3},
\label{Hamiltonian_matrix_form}
\end{eqnarray}
where the nearest neighbor ${\boldsymbol\delta}_{a}$ and next nearest neighbor vectors ${\boldsymbol\delta}'_{a}$ and the corresponding phase factors are
\begin{eqnarray}
&&{\boldsymbol\delta}_{1}=\left(\frac{1}{2},\frac{\sqrt{3}}{2}\right),\;\;\;
{\boldsymbol\delta}_{2}=\left(\frac{1}{2},-\frac{\sqrt{3}}{2}\right),\;\;\;
{\boldsymbol\delta}_{3}=\left(-1,0\right),
\nonumber \\
&&{\boldsymbol\delta}'_{1}=\left(-\frac{3}{2},-\frac{\sqrt{3}}{2}\right),\;\;\;
{\boldsymbol\delta}'_{2}=\left(\frac{3}{2},-\frac{\sqrt{3}}{2}\right),\;\;\;
{\boldsymbol\delta}'_{3}=\left(0,\sqrt{3}\right),
\nonumber \\
&&e_{a}=e^{i{\bf k}\cdot{\boldsymbol\delta}_{a}},\;\;\;
e'_{a}=2\cos{\bf k}\cdot{\boldsymbol\delta}'_{a},
\label{delta_definition}
\end{eqnarray}
These vectors and the resulting band structure using $t=2.8$eV$\equiv 1$ as the energy unit, together with $t'=0.1$eV$\equiv 0.036$ and $\lambda_{R}=0.56$eV$\equiv 0.2$ (a rather large RSOC just to demonstrate the splitting of bands), are shown in Fig.~\ref{fig:graphene_RSOC_band}. From the Hamiltonian matrix in Eq.~(\ref{Hamiltonian_matrix_form}), one can calculate the velocity operators by ${\hat j}_{\mu}({\bf k})=e\,\partial_{\mu}H({\bf k})$, which may then be used to calculate the optical conductivity and subsequently the quantum metric spectral function. 

From the form of the Hamiltonian in Eq.~(\ref{Hamiltonian_matrix_form}), one can immediately conclude that the next-nearest-neighbor hopping $t'$ does not affect the quantum metric and opacity of graphene, simply because it enters the diagonal element of the Hamiltonian in the form of an identity matrix $t'Z'\times I_{4\times 4}$. This means that $t'$ does give a $k$-dependent deformation to the dispersion without modifying the eigenstates, and the deformation is the same for all the bands at ${\bf k}$ and hence it does not affect the $\delta(\omega+\varepsilon_{\ell}^{\bf k}/\hbar-\varepsilon_{\ell '}^{\bf k}/\hbar)$ condition in the quantum metric spectral function either. As a result, the optical conductivity and opacity are not affected by $t'$.

\begin{figure}[ht]
\centering
\includegraphics[width=1.0\linewidth]{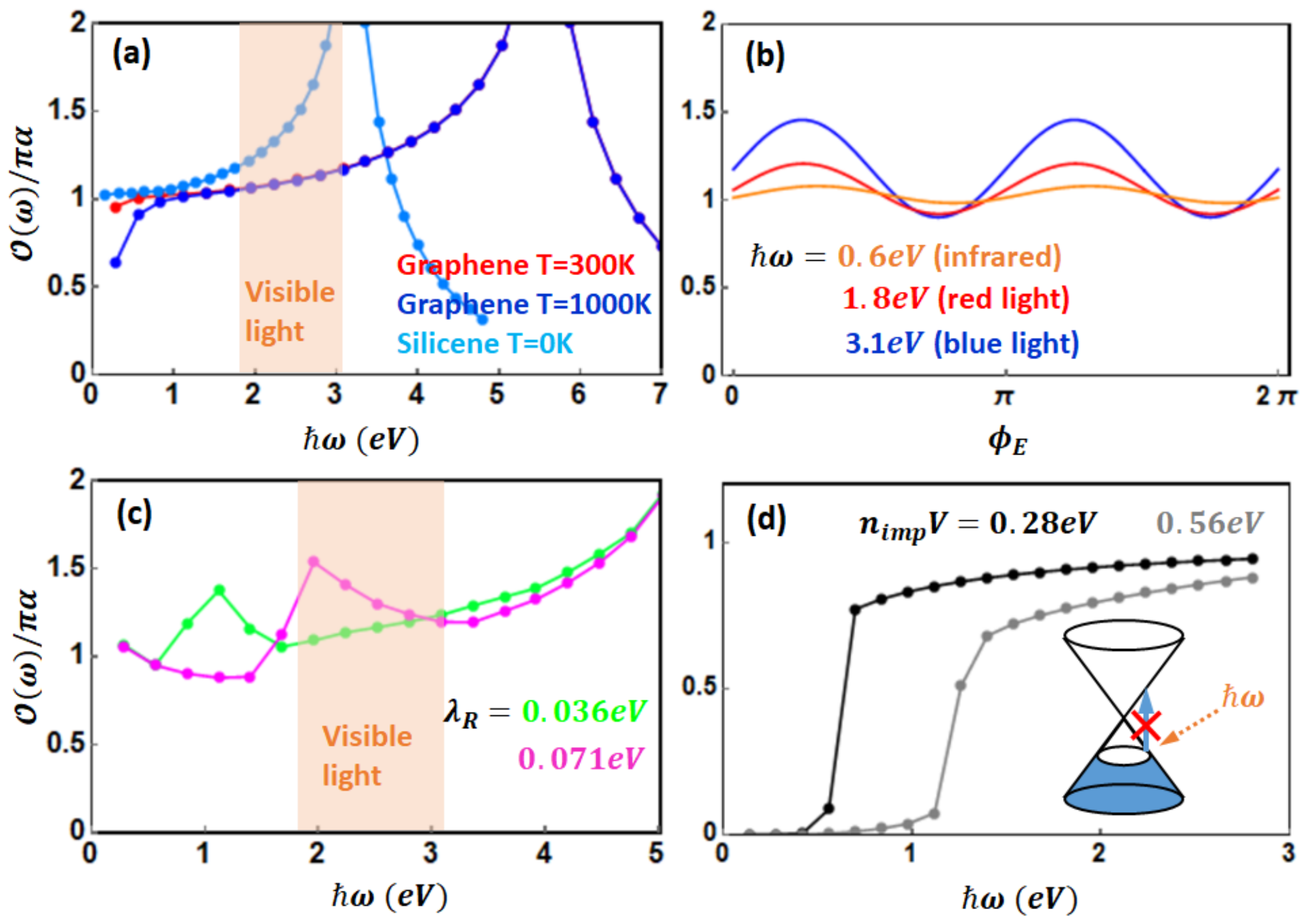}
\caption{(a) The opacity of graphene ${\cal O}(\omega)/\pi\alpha$ at $T=$ 300K, 1000K (a rather unrealistic temperature just to demonstrate the effect), and silicene at $T=0$ as a function of frequency of the unpolarized light. (b) The dependence of the opacity of graphene on the polarization angle $\phi_{E}$ at $T=0$. (c) The effect of RSOC on the opacity of graphene at $T=0$. (d) The effect of impurities on the opacity of linear Dirac model, with impurity density $n_{imp}$ and strength $V$. The inset shows schematically how the energy shift caused by the real part of self-energy blocks the optical transition. }
\label{fig:graphene_opacity_figure}
\end{figure}

However, the tight-binding band structure does feature an opacity that increases with frequency until the transition between van Hove singularities $\hbar\omega\approx 2t$, which makes the opacity in the visible light range to be $2.46\%<{\cal O}(\omega)<2.71\%$, as shown in Fig.~\ref{fig:graphene_opacity_figure} (a) and has already been investigated theoretically and observed experimentally\cite{Nair08,Stauber08,Kuzmenko08,Bruna09,Weber10,Nair10}. Nevertheless, since human eyes can hardly distinguish such a small deviation, the roughly constant opacity observed by human eyes against a light source of any color can still serve as a pedagogical example to demonstrate the topological charge. In fact, to our knowledge, this is the only known topological property of a material that can be directly perceived in the macroscopic scale, and is only recognized through the present work.

On the other hand, for 2D materials whose transition between van Hove singularities falls in the visible light range, such as silicene simulated by $t=1.6$eV\cite{Ezawa15} and shown in Fig.~\ref{fig:graphene_opacity_figure} (a), then the opacity will strongly depend on the color of the light. This result indicates that Dirac cone materials with larger hopping $t$, i.e., the linearity of Dirac cone extends beyond visible light range, are more ideal to visualize the topologically protected constant opacity. Nevertheless, because the opacity in the infrared region should still be $\pi\alpha$ for materials with a small $t$, the constant opacity may still be perceived by human eyes through an infrared lens. Finally, the finite temperature data in Fig.~\ref{fig:graphene_opacity_figure} (a) indicate that the thermal broadening in Eq.~(\ref{Ow_graphene_finiteT}) only reduces the opacity in the low frequency region $\hbar\omega\apprle k_{B}T$, which is far below and hence has negligible effect on the visible light range\cite{Stauber08,Kuzmenko08}.

\subsection{Polarization dependence and the effect of Rashba SOC on the opacity of graphene}

In this section, we demonstrate that the realistic band structure also influences the polarization dependence of the opacity. To calculate the dependence of the direction of polarization ${\hat{\boldsymbol\mu}}$ of the light, we consider ${\hat{\boldsymbol\mu}}$ to be pointing at the polar angle $\phi_{E}$ on the $xy$-plane of graphene. The electric field and the corresponding current operator in this situation may be decomposed into
\begin{eqnarray}
{\bf E}=E_{0}\left(\cos\phi_{E}{\hat{\bf x}}+\sin\phi_{E}{\hat{\bf y}}\right),\;\;\;
{\hat j}_{E}=\cos\phi_{E}{\hat j}_{x}+\sin\phi_{E}{\hat j}_{y}.
\nonumber \\
\end{eqnarray}
The usual linear response theory requires to calculate the correlator $\langle\left[{\hat j}_{E},{\hat j}_{E}\right]\rangle$, which yields an optical conductivity (suppressing $({\bf k},\omega)$)
\begin{eqnarray}
\sigma_{E}=\cos^{2}\phi_{E}\sigma_{xx}+\sin\phi_{E}\cos\phi_{E}\left(\sigma_{xy}+\sigma_{yx}\right)
+\sin^{2}\phi_{E}\sigma_{yy},
\nonumber \\
\end{eqnarray}
which is equivalent to calculating the quantum metric (note that $g_{xy}=g_{yx}$)
\begin{eqnarray}
g_{E}=\cos^{2}\phi_{E}g_{xx}+2\sin\phi_{E}\cos\phi_{E}g_{xy}
+\sin^{2}\phi_{E}g_{yy}.
\nonumber \\
\end{eqnarray}
The opacity will then carry the same angular dependence 
\begin{eqnarray}
{\cal O}_{E}(\omega,\phi_{E})&=&\cos^{2}\phi_{E}{\cal O}_{xx}(\omega)+2\sin\phi_{E}\cos\phi_{E}{\cal O}_{xy}(\omega)
\nonumber \\
&&+\sin^{2}\phi_{E}{\cal O}_{yy}(\omega).
\label{OwE_phiE}
\end{eqnarray}
where ${\cal O}_{\mu\nu}(\omega)$ corresponds to the contribution coming from $g_{\mu\nu}$. Note that if a light source is not polarized, then averaging over the angle yields an opacity ${\overline{\cal O}}(\omega)=\left[{\cal O}_{xx}(\omega)+{\cal O}_{yy}(\omega)\right]/2$, meaning that the contribution from the off-diagonal element of the quantum metric $g_{xy}$ drops out. However, for a polarized light, because ${\cal O}_{xx}(\omega)\approx {\cal O}_{yy}(\omega)$ (for instance, $({\cal O}_{xx}(\omega),{\cal O}_{yy}(\omega),{\cal O}_{xy}(\omega))=(1.175,1.182,0.277)\times\pi\alpha$ for blue light $\hbar\omega=3.1$eV), the variation of ${\cal O}_{E}(\omega,\phi_{E})$ in Eq.~(\ref{OwE_phiE}) as a function of the polarization angle $\phi_{E}$ mainly comes from the $2\sin\phi_{E}\cos\phi_{E}{\cal O}_{xy}(\omega)$ term contributed from the optical Hall conductance $\sigma_{xy}(\omega)$, which is originated from the off-diagonal element of the quantum metric $g_{xy}$.

Our calculation reveals a $15\%$ to $30\%$ variation with the polarization angle $\phi_{E}$ in the visible light range, as shown in Fig.~\ref{fig:graphene_opacity_figure} (b). Since this variation is mainly contributed from the optical Hall conductance $\sigma_{xy}(\omega)$ originated from $g_{xy}$, one can actually estimate the magnitude of $\sigma_{xy}(\omega)$ by simply rotating a graphene sheet against a polarized light and seeing how its opacity varies. However, this also indicates that it is hard to associate the opacity against a polarized light to the topological charge since it is not a constant of $\phi_{E}$, and suggesting that only unpolarized light that averages over $\phi_{E}$ can serve this purpose. 


Furthermore, for 2D materials that are not perfectly flat but have some buckling structure, the breaking of inversion symmetry can induce a Rashba spin-orbit coupling (RSOC). By incorporating the RSOC of strength $\lambda_{R}$ into the honeycomb lattice\cite{deSousa21_RSOC_graphene} as described by Eq.~(\ref{Hamiltonian_graphene_Rashba_mag}), we obtain an opacity that shows an anomaly at frequency $\omega\sim 3\lambda_{R}$ owing to the splitting of the bands, as shown in Fig.~\ref{fig:graphene_opacity_figure} (c). Thus if the RSOC of some material happens to fall in the visible light region $1.8$eV$\apprle 3\lambda_{R}\apprle 3.1$eV, then its opacity may strongly depend on the color of the light. However, such a strong RSOC seems rather unlikely, since most of known 2D materials have an RSOC of the order of 10meV\cite{Liu11_2D_Hamiltonian}, and hence its influence on the visible light range is negligible.

\subsection{Influence of impurity scattering on the opacity of graphene}

Concerning various sources of scattering, graphene under the influence of electron-electron interaction has been investigated intensively\cite{Herbut08,Elias11}, which only yields a very small correction to the opacity\cite{Sheehy09}. In contrast, we consider the effect of random point-like impurities\cite{Ando02}, whose effect is detailed in Appendix \ref{apx:impurity_effect}. The results shown in Fig.~\ref{fig:graphene_opacity_figure} (d) indicate an opacity that is strongly suppressed in the low frequency regime $\hbar\omega\apprle 2n_{imp}V$, which is caused by the real part of the self-energy that acts like a chemical potential, rendering the valence band states with energy less than $\varepsilon_{n}^{\bf k}<n_{imp}V$ empty. As a result, the optical absorption is strongly suppressed and hence the material actually becomes more transparent at $\hbar\omega\apprle 2n_{imp}V$, as indicated by the inset of Fig.~\ref{fig:graphene_opacity_figure} (d). This suppression is in qualitative agreement with the experiment in fluorinated graphene, where a reduced low frequency opacity is observed, and the reduction region can be extended to the visible light range by increasing fluorine concentration\cite{Nair10}. In fact, changing the chemical potential should also cause such a suppression, as has been observed experimentally\cite{Li08}. Thus to see the topological charge from the opacity in the visible light range, a clean and unbiased graphene is needed. Finally, we remark that the opacity is frequency-independent only if the topological material is gapless, as we demonstrate in the Appendix \ref{apx:opacity_Chern} using 2D Chern insulator as a counterexample. In addition, in Appendix \ref{apx:extracting_fine_structure_constant} we also discuss the possibility of extracting fine-structure constant accurately from frequency-dependence of the opacity.

\subsection{Opacity of 3D TIs} 

The bulk of 3D TI like Bi$_{2}$Se$_{3}$ and Sb$_{2}$Te$_{3}$ usually have a direct band gap $2M\sim 0.5$eV\cite{Zhang09,Liu10}, so one may expect it to be transparent in the infrared region. However, note that thin films of TIs have surface states in the top and bottom surfaces where the light passes through, and each has two spin species, hence one may label them by the four spin-surface flavors $\gamma=({\rm top}\uparrow,{\rm top}\downarrow,{\rm bottom}\uparrow,{\rm bottom}\downarrow)$. Each $\gamma$ is described by one of the Hamiltonians in Eq.~(\ref{graphene_linear_Dirac_Hamiltonian}) with a proper assignment of $k_{x}$ and $k_{y}$\cite{Zhang09,Liu10,Qi11}, yielding the same opacity $\pi\alpha$ as graphene, and is also independent of the frequency and polarization in the infrared region owing to the topological charge of the surface state, equivalently the bulk topological invariant due to the bulk-edge correspondence. Remarkably, this implies that one can literally see the topological surface states by naked eyes from the opacity through an infrared lens. Moreover, the opacity should be independent of the thickness of the TI provided the material is thicker than the decay length of surface states (same as the correlation length $\hbar v_{F}/M\sim$ nm\cite{Chen17,Chen19_universality}), and should be the same for all 3D TIs regardless of many details such as the Fermi velocity, lattice constant, and chemical composition, which is ready to be verified by naked eyes.

We remark that several experiments have already hinted this constant opacity of 3D TI thin films. Peng et al.~measured the Bi$_{2}$Se$_{3}$ thin films deposit on the transparent substrate of mica\cite{Peng12}, which reveals that the transmittance at different thickness all seem to saturate to the same value in the infrared region, although the Fabry–-Perot interference hinders the extraction of a precise value at the saturation. Chuai et al.~measured the Bi$_{2}$Se$_{3}$ thin film deposit on p-Si (111) substrate\cite{Chuai23}, and showed that the transmittance of the film is higher than $90\%$ and is a constant of frequency in a wide range of infrared region, although the precise value has not been quantified. Despite further efforts are required to compare experiments with our theory, these preliminary results of transmittance that are roughly constant of frequency and thickness seem to be highly encouraging.


\subsection{Optical absorption of 3D Dirac and Weyl semimetals} 

We proceed to consider the low energy sector of type-I 3D Weyl semimetals, such as TaAs and TaP\cite{Weng15,Huang15}, which contain pairs of Weyl points\cite{Yan17,Armitage18,Lv21}. Their low energy eigenstates can be labeled by the valley index $\gamma=\left\{1,2...N_{W}\right\}$, where $N_{W}$ is the number of Weyl points. The Hamiltonian in the infrared region for each $\gamma$ is well described by the Dirac model 
\begin{eqnarray}
H^{\gamma}({\bf k})=\pm\left(vk_{x}\sigma_{x}+vk_{y}\sigma_{y}+vk_{z}\sigma_{z}\right)=\pm{\bf d}\cdot{\boldsymbol\sigma},\;\;\;
\label{Weyl_Hamiltonian}
\end{eqnarray}
and we assume a cubic lattice of unit cell volume $V_{D}=a^{3}$ for simplicity. The topological charge of each Weyl point is calculated by the integration of Berry curvature over a spherical surface of any radius that encloses the Weyl point, whose sign depends on the chirality
\begin{eqnarray}
&&{\cal C}=\pm\frac{1}{4\pi}\int d\phi\int d\theta\frac{1}{d^{3}}\varepsilon^{ijk}d_{i}\partial_{\theta}d_{j}\partial_{\phi}d_{k}
\nonumber \\
&&=\pm\frac{1}{4\pi}\int d\phi\int d\theta\sin\theta=\pm 1,
\label{topo_charge_Weyl}
\end{eqnarray}
and we denote the integrand by $J_{\bf k}=\varepsilon^{ijk}d_{i}\partial_{\theta}d_{j}\partial_{\phi}d_{k}/d^{3}=\sin\theta$. On the other hand, the quantum metric of each spin-valley flavor $\gamma$ on the spherical surface is 
\begin{eqnarray}
g_{\theta\theta}^{\gamma}=\frac{1}{4},\;\;\;g_{\phi\phi}^{\gamma}=\frac{1}{4}\sin^{2}\theta,\;\;\;
g_{\theta\phi}^{\gamma}=g_{\phi\theta}^{\gamma}=0,
\end{eqnarray}
which satisfies the metric-curvature correspondence 
\begin{eqnarray}
\sqrt{\det g^{\gamma}}=\frac{1}{4}|J_{\bf k}|.
\end{eqnarray}
In the Cartesian coordinates, the metric is given by
\begin{eqnarray}
g_{\mu\mu}^{\gamma}=\frac{1}{4k^{4}}\left(k^{2}-k_{\mu}^{2}\right),\;\;\;
g_{\mu\nu}^{\gamma}|_{\mu\neq\nu}=-\frac{k_{\mu}k_{\nu}}{4k^{4}}.
\end{eqnarray}
and hence the trace of the quantum metric at ${\bf k}$ for each spin-valley flavor $g_{xx}^{\gamma}+g_{yy}^{\gamma}+g_{zz}^{\gamma}=1/2k^{2}$ only depends on the module $k$ but not the direction ${\hat{\bf k}}$ of the momentum. As a result, using the definition of $g_{\mu\mu}^{\gamma}({\bf k},\omega)$ in Eq.~(\ref{fidelity_number_spec_fn}) with $D=3$, the conductivity in real space summing over three crystalline directions is 
\begin{eqnarray}
&&\sum_{\mu=x,y,z}\sigma_{\mu\mu}(\omega)
=\frac{N_{W}e^{2}\omega}{4\pi\hbar^{2}}\left[\frac{1}{4\pi}\int d\phi\int d\theta\sin\theta\right]
\nonumber \\
&&\times\int dk\left[f\left(-vk\right)-f\left(vk\right)\right]\frac{\hbar}{2v}\delta(k-\frac{\hbar\omega}{2v})
\nonumber \\
&&=\frac{N_{W}e^{2}\omega|{\cal C}|}{8\pi\hbar v}\left[f\left(-\frac{\hbar\omega}{2}\right)-f\left(\frac{\hbar\omega}{2}\right)\right],
\label{sigma_Weyl_sumxyz}
\end{eqnarray}
since the bracket $\left[...\right]$ in the second line is precisely the topological charge $|{\cal C}|$ in Eq.~(\ref{topo_charge_Weyl}). The absorption power summing over the three crystalline directions and then divided by the incident power per unit cell volume $W_{i}=c\varepsilon_{0}E_{0}^{2}/2a$ yields
\begin{eqnarray}
&&\sum_{\mu=x,y,z}\frac{W_{a}^{\mu}(\omega)}{W_{i}}
\nonumber \\
&&=\alpha|{\cal C}|\left(\frac{N_{W}a}{2v}\right)\omega\left[f\left(-\frac{\hbar\omega}{2}\right)-f\left(\frac{\hbar\omega}{2}\right)\right].
\label{Weyl_absorption_power}
\end{eqnarray}
whose zero temperature limit is linear in frequency, as has been pointed out theoretically\cite{Hosur12,Bacsi13,Timusk13,Ashby14} and experimentally observed\cite{Xu16}. Our result further suggests that the optical absorption power summing over three crystalline directions is directly proportional to the module of the topological charge $|{\cal C}|$ and fine-structure constant $\alpha$. In addition, since it is proportional to the module, even if two Weyl nodes of opposite chirality merge together to form a Dirac semimetal\cite{Young12_Dirac_semimetal,Manes12}, the absorption power is still that described by Eq.~(\ref{Weyl_absorption_power}), whose linearity in frequency has been observed experimentally\cite{Chen15_Dirac_optical}. Physically, this means that Dirac and Weyl semimetals appear darker under higher frequency light in the infrared region, which should be detectable by human eyes through an infrared lens. Finally, we remark that 3D topological semimetals have surface states too, but since the bulk already absorbs light, the contribution from the surface states should be negligible in the bulk limit.



\section{Conclusions} 

In summary, we clarify that the approximate frequency-independence of the opacity of disorder-free and unbiased graphene can be attributed to the topological charge of Dirac points. In other words, the roughly $\pi\alpha\approx 2.3\%$ opacity of graphene against any sources of unpolarized visible light is a manifestation of topological charge. The same analysis applied to 3D TIs suggests that their opacity in the infrared region ideally is also the bulk topological invariant times $\pi\alpha$ independent of frequency, polarization, and thickness of the material. In contrast, for 3D Weyl and Dirac semimetals, the fine-structure constant and topological charge also determine the linear dependence of optical absorption power on frequency. Our results indicate that topological charges and surface states can be directly perceived by naked eyes in the macroscopic scale, and it offers a very accessible way to estimate the fine-structure constant in a topologically protected manner.



\appendix

\section{Effect of impurities \label{apx:impurity_effect}}

\begin{figure}[ht]
\centering
\includegraphics[width=0.99\linewidth]{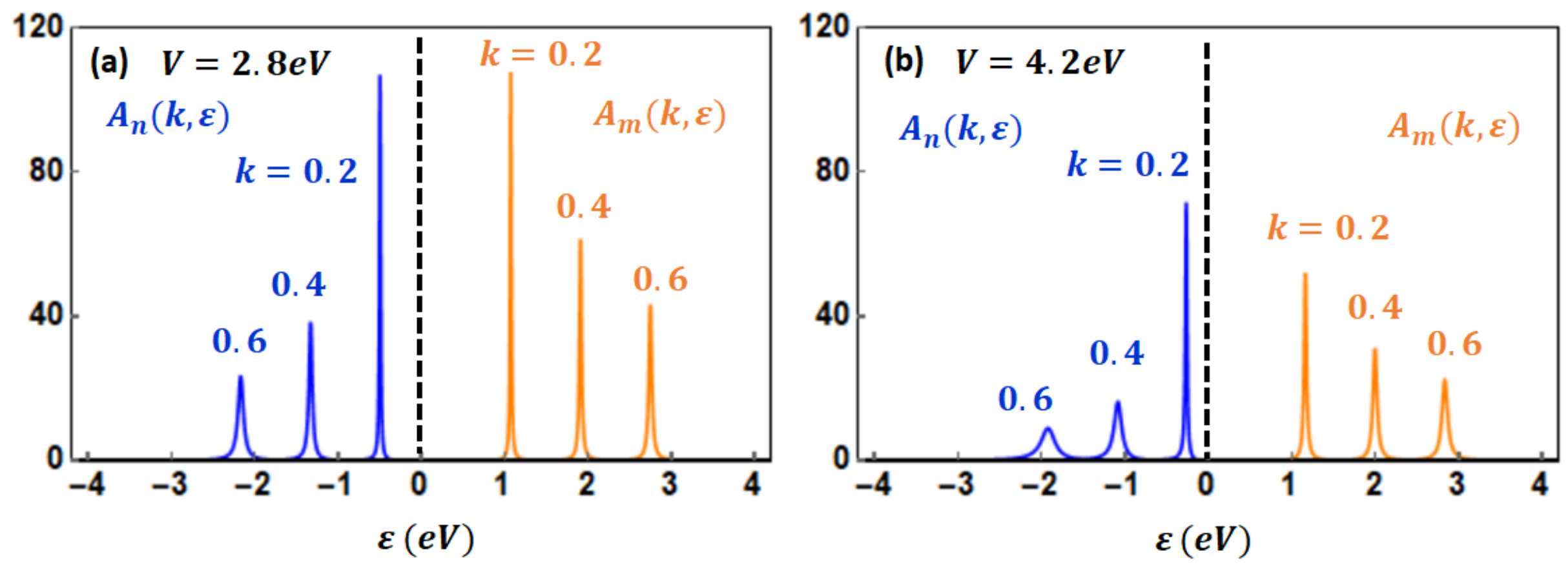}
\caption{Single-particle spectral function for the valence band $A_{n}({\bf k},\omega)$ and the conduction band $A_{m}({\bf k},\omega)$ of disordered graphene, where we fix the impurity density at $n_{imp}=10\%$ and plot the results at impurity potential (a) $V=0.28$eV and (b) $V=4.2$eV. The dashed line labels the Dirac point of the pristine graphene. The result suggests a shift of energy spectrum towards positive energy caused by the impurities. }
\label{fig:Akw_graphene_imp}
\end{figure}

We proceed to utilize a many-body formalism to investigate the opacity of graphene under the influence of impurities. To give an analytical result at low energy, we adopt the spinless linear Dirac model of the basis $(c_{A{\bf k}},c_{B{\bf k}})$ near the ${\bf K}$ point as described by Eq.~(\ref{graphene_linear_Dirac_Hamiltonian}), and consider the potential scattering
\begin{eqnarray}
{\hat V}=V\left(\begin{array}{cc}
1 & 0 \\
0 & 1 
\end{array}\right),
\end{eqnarray}
where $V$ is the strength of impurity potential. We will focus on only one spin species near the ${\bf K}$ valley, since all the four spin-valley flavors give the same result. The corresponding vertex for the $\ell=\left\{n,m\right\}$ band is
\begin{eqnarray}
V_{\bf kk'}=\langle \ell{\bf k}|{\hat V}|\ell{\bf k'}\rangle=\langle n\phi|{\hat V}|n\phi '\rangle
=\frac{V}{2}\left[1+e^{i(\phi '-\phi)}\right],
\label{impurity_vertex}
\nonumber \\
\end{eqnarray}
yielding the $T$-matrix for the $\ell$ band
\begin{eqnarray}
&&T_{\bf kk'}^{\ell}(\omega)=\frac{V}{2}\left[1+e^{i(\phi '-\phi)}\right]\sum_{s=0}^{\infty}\left(\frac{V}{2}\int\frac{k_{1}dk_{1}}{2\pi/a^{2}}G_{\ell}({\bf k_{1}},\omega)\right)^{s}
\nonumber \\
&&=\frac{\frac{V}{2}\left[1+e^{i(\phi '-\phi)}\right]}{1-\frac{V}{2}\int\frac{k_{1}dk_{1}}{2\pi/a^{2}}G_{\ell}({\bf k_{1}},\omega)}.
\end{eqnarray}
So we are lead to the integration of the retarded Green's function over the module of the momentum, assuming that it has a cut-off $0<k_{1}<\pi/a$
\begin{widetext}
\begin{eqnarray}
\int_{0}^{\pi/a}\frac{k_{1}dk_{1}}{2\pi/a^{2}}G_{\ell}({\bf k_{1}},\omega)
=\int_{0}^{\pi/a}\frac{k_{1}dk_{1}}{2\pi/a^{2}}\left[\frac{1}{\omega\pm vk}-i\pi\delta(\omega\pm vk)\right],
\end{eqnarray} 
where $+$ is for the valence band $\ell=n$ and $-$ is for the conduction band $\ell=m$, which leads to the result
\begin{eqnarray}
&&T_{\bf kk'}^{n}(\omega)=\frac{\frac{V}{2}\left[1+e^{i(\phi-\phi ')}\right]}{1-\frac{Va^{2}}{4\pi v^{2}}\left[\frac{\pi v}{a}-\omega\ln\left|\frac{\pi v/a+\omega}{\omega}\right|\right]-i\frac{Va^{2}\omega}{4v^{2}}\theta(0>\omega>-\pi v/a)},
\nonumber \\
&&T_{\bf kk'}^{m}(\omega)=\frac{\frac{V}{2}\left[1+e^{i(\phi-\phi ')}\right]}{1-\frac{Va^{2}}{4\pi v^{2}}\left[-\frac{\pi v}{a}-\omega\ln\left|\frac{\pi v/a-\omega}{\omega}\right|\right]+i\frac{Va^{2}\omega}{4v^{2}}\theta(0<\omega<\pi v/a)},
\end{eqnarray}
where the $\theta$-function ensures the range of the frequency $\omega$. The self-energy at momentum ${\bf k}$ is simply the $T$-matrix at ${\bf k=k'}$ multiplied by the density of the impurity $\Sigma_{\ell}({\bf k},\omega)=n_{imp}T_{\bf kk}^{\ell}(\omega)$, which can be used to calculate the spectral function by
\begin{eqnarray}
A_{\ell}({\bf k},\omega)=-\frac{1}{\pi}{\rm Im}\,G_{\ell}({\bf k},\omega)=-\frac{1}{\pi}\frac{{\rm Im}\Sigma_{\ell}({\bf k},\omega)}{(\omega\pm v k-{\rm Re}\Sigma_{\ell}({\bf k},\omega))^{2}+{\rm Im}\Sigma_{\ell}({\bf k},\omega)^2}.
\end{eqnarray}
After the spectral function is obtained, we further use the full Green's function approximation to obtain the dressed quantum metric spectral function $\tilde{g}_{\mu\nu}^{\gamma}({\bf k},\omega)$ for the spin-valley flavor $\gamma$ by
\begin{eqnarray}
\tilde{g}_{\mu\nu}^{\gamma}({\bf k},\omega)=\hbar g_{\mu\nu}^{\gamma}\int_{-\infty}^{\infty}d\varepsilon\,A_{n}({\bf k},\varepsilon)A_{m}({\bf k},\varepsilon+\hbar\omega)\left[f(\varepsilon)-f(\varepsilon+\hbar\omega)\right]
\end{eqnarray}
\end{widetext}
where $g_{\mu\nu}^{\gamma}$ is the unperturbed quantum metric. This dressed spectral function can then be used to calculate the opacity using the formalism presented in Sec.~\ref{sec:relating_absorption_to_quantum_geometry} and \ref{sec:opacity_pristine_graphene}. The spectral function $A_{\ell}({\bf k},\omega)$ shown in Fig.~\ref{fig:Akw_graphene_imp} exhibits a broadening of quasilarticle peak that gradually diminishes as $k\rightarrow 0$ owing to the imaginary part of self-energy, and a positive energy shift relative to the Dirac point of the pristine graphene due to the real part roughly given by the density of impurities times their strength $n_{imp}V$. This positive energy shift strongly suppresses the optical conductivity in the low frequency regime $\hbar\omega\apprle 2n_{imp}V$, as argued in Sec.~\ref{sec:realistic_factors_opacity}. 

\section{Opacity of a Chern insulator \label{apx:opacity_Chern}}

We remark that the constant opacity $\pi\alpha$ seems to manifest only if the unperturbed topological material is gapless. As a counter example, consider a 2D Chern insulator described by adding a mass term $M\sigma_{z}$ into the linear Dirac model in Eq.~(\ref{graphene_linear_Dirac_Hamiltonian}), which gaps out the Dirac cone. The resulting fidelity number spectral function is
\begin{eqnarray}
&&{\cal G}_{\mu\mu}(\omega)
\nonumber \\
&&=\left[\frac{1}{16\pi\omega}+\frac{M^{2}}{4\pi\hbar^{2}\omega^{3}}\right]\left[f\left(-\frac{\hbar\omega}{2}\right)-f\left(\frac{\hbar\omega}{2}\right)\right]_{\omega\geq 2|M|/\hbar},
\nonumber \\
\end{eqnarray}
which yields an opacity that depends on the band gap $M$ and is not a constant of frequency. Since the opacity depends on the color of the light and is not directly proportional to the Chern number, it is hard to argue that one can see the bulk topological invariant by naked eyes in this case.

\section{Possibility of extracting fine-structure constant accurately from the opacity \label{apx:extracting_fine_structure_constant}}

Given all these complications in reality, a question that naturally arises is whether it is possible to extract the fine-structure constant $\alpha$ up to very high precision from the opacity. We anticipate that this is possible by measuring the opacity at zero temperature in the low frequency limit against an unpolarized light source, where it has been pointed out in Ref.~\onlinecite{Stauber08} that within tight-binding model, the opacity should depend on frequency quadratically 
\begin{eqnarray}
\lim_{\left\{T,\omega\right\}\rightarrow 0}{\cal O}(\omega)=\pi\alpha+\beta\omega^{2},
\label{alpha_wsquare}
\end{eqnarray}
where $\beta$ is a nonuniversal coefficient that depends on the Fermi velocity of the material. Provided that the sample is clean enough and no RSOC is present, $\alpha$ extracted by fitting the experimental data by Eq.~(\ref{alpha_wsquare}) should be very accurate in all the 2D Dirac semimetals mentioned above. The feasibility of this frequency-dependence measurement awaits to be verified.

\bibliography{Literatur}

\end{document}